\documentclass[fleqn,10pt]{wlscirep}
\usepackage[utf8]{inputenc}
\usepackage[T1]{fontenc}
\usepackage[english]{babel}
\usepackage{subcaption}
\usepackage{amsmath,amsfonts,amssymb}
\usepackage{graphicx}
\usepackage{xcolor}
\usepackage{soul}
\usepackage{physics}
\sethlcolor{yellow}
\usepackage[colorlinks=true, allcolors=blue]{hyperref}
\usepackage{appendix}
\usepackage{tcolorbox}

\title{Quantum Hilbert Transform}

\author[1,*]{Nitin Jha}
\author[1]{Abhishek Parakh}
\affil[1]{Kennesaw State University, Marietta, GA, USA}

\affil[*]{Corresponding author: njha1@students.kennesaw.edu}


\begin{abstract}
The Hilbert transform has been one of the foundational transforms in signal processing, finding it's way into multiple disciplines from cryptography to biomedical sciences. However, there does not exist any quantum analogue for the Hilbert transform. In this work, we introduce a formulation for the quantum Hilbert transform (QHT)and apply it to a quantum steganography protocol. By bridging classical phase-shift techniques with quantum operations, QHT opens new pathways in quantum signal processing, communications, sensing, and secure information hiding.

\end{abstract}
\begin{document}

\flushbottom
\maketitle
%
%
\thispagestyle{empty}

\section{Introduction}

The Hilbert transform is one of the most significant signal and image processing tools used to generate an analytic signal from a real-valued signal by introducing a $\pm\pi/2$ phase shift to its frequency components. This transformation has been of immense importance in cases where one needs to extract the instantaneous phase, frequency, or amplitude of any given signal crucial in applications ranging from envelope detection in biomedical signals to feature extraction in radar systems \cite{Hilbert2006, Hahn1996HilbertTI, oppenheim2021discrete, Feldman2008TheoreticalAA,khare2022vhers}. Hilbert transform enables the use of Single Sideband modulation (SSB) which reduces the bandwidth requirements and thus has even been exploited to embed imperceptible covert channels in audio streams  \cite{Li2011, HilbertHiding2006} . The use of Hilbert transform in steganography is often done by using Hilbert-Huang transform. There have been several studies which uses the classical Discrete Hilbert transform (DHT) and its adaptive extension, the Hilbert–Huang transform (HHT), to hide and detect secret information in a variety of signal and image domains \cite{Hilberthuang, kandregula2009, wu2011detection, sharma2022Hilbert}.

To exploit the usefulness of Hilbert transform in digital domain, several discrete Hilbert transform (DHT) algorithms have been developed. Kak’s lifting‐based DHT \cite{Kak1970DHT} and subsequent trigonometric variants\cite{pei2002computation, kak1977discrete, DHT2}  allow efficient computation via the discrete Fourier transform (DFT). Hilbert-Huang transform  is an adaptive, two-stage signal analysis technique that first decomposes any nonlinear, nonstationary time series into a finite set of data-driven Intrinsic Mode Functions (IMFs) via Empirical Mode Decomposition. Hilbert transform is then applied to determine instantaneous amplitude and frequency information. This gives us \textit{Hilbert Spectrum}.  Kandregula\cite{kandregula2009} showed that a finite-length discrete Hilbert transform can carry binary data by alternately transmitting the original or a subtly phase-shifted version of the host signal—an approach that leverages our near-complete insensitivity to small phase changes to embed information imperceptibly and reversibly in audio and other one-dimensional signals. Tan et al. \cite{tan2007steganalysis} extended this idea to \textit{Enhanced Bit-Plane Complexity Segmentation (BPCS) steganography} by applying empirical mode decomposition and the Hilbert transform to bit-plane complexity difference sequences; the resulting phase-sequence feature vectors, when classified with an Support Vector Machine (SVM), reliably expose hidden payloads in both spatial and JPEG2000 images. Li and Zhou \cite{wu2011detection}, employed the Hilbert–Huang transform to break an image’s pixel-value histogram into intrinsic mode functions and then extracted instantaneous amplitude and frequency descriptors—features that, fed into a machine-learning classifier, achieve very high accuracy in detecting embedded data across diverse image formats. 

Subsequently, in the field of quantum computing, the quantum Fourier transform (QFT) has emerged as a core primitive for phase estimation, period finding, and quantum signal processing \cite{shor1999polynomial, zhouquant2017}.  Quantum signal processing is a rapidly growing area of research and several quantum analogues of classical transforms, such as the wavelet and cosine transforms, have been proposed \cite{pang2019signal, yin2021quantum, motlagh2024generalized}. Yet, despite the importance of both the classical Hilbert transform and the QFT in their respective domains, no direct quantum analogue of the Hilbert transform is proposed at this point. A quantum Hilbert transform (QHT) would enable coherent, unitary phase‐shift operations on multiqubit states, opening new possibilities in quantum signal analysis, quantum communications, and quantum steganography.

This paper develops a quantum analogue of the classical Hilbert transform and introduces its application to quantum steganography. The paper is structured as follows. Section \ref{HT} gives the mathematical definition of the continuous and discrete Hilbert transforms. We also explain the basic mathematical formulation of quantum Fourier transform, which is central to the development of the proposed quantum Hilbert transform. Section \ref{QHT} defines the quantum Hilbert transform, analogous to the classical Hilbert transform. This study defines a discrete quantum Hilbert transform based on a discrete quantum Fourier transform. Section \ref{sec:examples} provides example use cases for the newly proposed QHT. Finally, Section {\ref{sec:discussion}} provides a discussion and introduces the applications of QHT in steganography, outlining the importance of this work.

\subsection{Hilbert transform}
\label{HT}

The Hilbert transform ($\mathcal{H}[g(t)]$), for a signal $g(t)$, is defined as follows, \cite{Hilbert2006}
\begin{equation}
    \mathcal{H}[g(t)] = g(t) * \frac{1}{\pi t} = \int_{-\infty}^{\infty}\frac{g(t-\tau)}{\tau}\mathrm{d}\tau,
    \label{h(g(t))}
\end{equation}
where the Hilbert transform is the convolution product of the signal with $1/\pi t$. The Hilbert transform is often denoted as $\hat{g}(t)$ or $[g(t)]\ \hat{}$. This particular definition of HT is rather complicated at first glance due to the presence of improper integrals containing singularities. Hilbert transform is defined as \textit{Cauchy's principal value} of the integral in eq(\ref{h(g(t))}). This is defined as, 
\begin{equation}
    \mathcal{H}[g(t)] = \frac{1}{\pi}\  \lim_{\epsilon\to 0^+}\left(\int_{t-1/\epsilon}^{t-\epsilon} \frac{g(\tau)}{t-\tau}\mathrm{d} \tau + \int_{t+\epsilon}^{t+1/\epsilon} \frac{g(\tau)}{t-\tau}\mathrm{d} \tau \right).
\end{equation}
Therefore, the integral from eq(\ref{h(g(t))}) represents the Cauchy principal value \cite{Hilbert2006}. To understand HT in an easier way, we look at it's interaction with Fourier transform. Thus, we move from the time to frequency domain.

For the signal $1/(\pi t)$, the Fourier transform can be written as, 
\begin{equation}
    - j \, \text{sgn}(f) = 
\begin{cases} 
      -j, & \text{if } f > 0 \\ 
      0, & \text{if } f = 0 \\ 
      j, & \text{if } f < 0 
\end{cases}
\label{FT-sgn}
\end{equation}
Therefore, if $g(t)$ signal has a Fourier transform $G(f)$, the using the convolution property of FT, we can write $\mathcal{H}(g(t))$ in frequency domain as, 
\begin{equation}
    \hat{G}(f) = -j\ \text{sgn}(f)G(f).
    \label{FT-HT}
\end{equation}
Its evident from eq(\ref{FT-HT}) that HT doesn't change the magnitude of the input signal, rather than changes the phase of the signal. All the positive frequency components are multiplied by $-j$ (phase shift of $-\pi/2$), and negative frequency components are multiplied by $j$ (phase shift of $\pi/2$). Therefore, HT exchanges the real and imaginary parts of $G(f)$ (changes sign of one of the components) in the Fourier Basis.

\subsubsection{Discrete Hilbert transform}
\label{SEC:DHT}
The Discrete Hilbert transform (DHT) of a real-valued sequence $x[n]$ can be defined in the frequency domain using the Discrete Fourier transform (DFT). We start with computing the DFT of the real-valued sequence. Multiply the DFT coefficients by a filter to shift the negative frequency components to zero and double the positive frequency components. This operation corresponds to the Hilbert transform in the frequency domain. We compute the DFT as,

\begin{equation}
    X[k] \;=\;\sum_{n=0}^{N-1}x[n]\,e^{-2\pi i\,k n/N}.
\end{equation}
Now, we need to define the filter $H[k]$ is defined as,
\begin{equation}
    H[k] = 
\begin{cases} 
1, & \text{if } k = 0 \\
2, & \text{if } 1 \leq k \leq \frac{N}{2} - 1 \\
1, & \text{if } k = \frac{N}{2} \text{ (if N is even)} \\
0, & \text{if } \frac{N}{2} + 1 \leq k \leq N-1 
\end{cases}
\label{Hilbert-filter}
\end{equation}
where \( N \) is the length of the sequence. Now we construct the analytic signal as follows,

\begin{equation}
     \widetilde X[k] \;=\; X[k]\,H_{\mathrm{an}}[k].
\end{equation}
Take the inverse DFT,
\begin{equation}
         x_{\rm an}[n]
     \;=\;
     \frac{1}{N}
     \sum_{k=0}^{N-1}
     \widetilde X[k]\,e^{+2\pi i\,k n/N}.
\end{equation}
And, finally, we can get the DHT as, (by extracting the imaginary part of the analytical signal)
\begin{equation}
     \hat{x}[n]
     \;=\;
     \Im\bigl\{x_{\rm an}[n]\bigr\}.
    \label{final-dht}
\end{equation}

\subsection{quantum Fourier transform}
Discrete Fourier transform (DFT) is defined to take a unit vector, $\hat{x} = (x_0, x_1\ldots x_{N-1}) \in \mathbb{C}^N$, produce another unit vector, $\hat{y} = (y_0, y_1\ldots y_{N-1})$. Where the output unit vector, $\hat{y}_k$, can be written as, \cite{zhouquant2017}
\begin{equation}
    y_k = \frac{1}{\sqrt{N}}\sum^{N-1}_{j=0}e^{2\pi ijk/N}x_j,\ k=0,1,\ldots, N-1.
    \label{dft}
\end{equation}
The quantum Fourier transform (QFT) performs an operation similar to the discrete Fourier transform but in amplitudes. Mathematically, 

\begin{equation}
    \sum^{N-1}_{j=0}x_j|j\rangle \to \sum^{N-1}_{k=0}y_k|k\rangle.
    \label{QFT}
\end{equation}

From eq(\ref{QFT}), we can see that QFT takes a quantum state from a computational basis to Fourier basis. The state information is encoded in Fourier coefficients.
\begin{equation}
    \text{QFT}(|\psi\rangle) = \frac{1}{\sqrt{N}}\sum_{j=0}^{N-1}\left(\sum^{N-1}_{k=0}x_ke^{2\pi ijk/N} \right)|j\rangle,
    \label{qft-fourier}
\end{equation}
where $\beta_j = (\sum^{N-1}_{k=0}x_ke^{2\pi ijk/N})$ are the Fourier coefficients. We will use the phase information extracted from these Fourier coefficients in the next section when we define the quantum Hilbert transform. \\

\section{quantum Hilbert transform}
\label{QHT}
Now that we have established a relationship between the Hilbert transform and the Fourier transforms, we can define the quantum analogue of the classical Hilbert transform, which we shall refer to as the quantum Hilbert transform (QHT). The QHT might serve an essential tool in quantum signal processing, mirroring the role of the classical Hilbert transform in the analysis and filtering of signals.
\begin{enumerate}
    \item To begin, let us first define a quantum state. For a system with \(N\) possible states, the general form of a quantum state $|\psi\rangle$ can be expressed as:

\begin{equation}
    |\psi\rangle = \sum^{N-1}_{x=0} \alpha_x |x\rangle,
    \label{quant-state}
\end{equation}

where $\alpha_x$ are complex coefficients that determine the amplitude and phase of each basis state $|x\rangle$. The coefficients $\alpha_x $ are subject to the normalization condition $\sum_x |\alpha_x|^2 = 1$.

    \item With the quantum state  $|\psi\rangle$ defined, we now apply the quantum Fourier transform (QFT) to this state. When the QFT is applied to the quantum state  $|\psi\rangle$, the resulting state $|\phi\rangle$ can be written as,

\begin{equation}
    |\phi\rangle = \text{QFT}(|\psi\rangle) = \sum^{N-1}_{j=0} \beta_j |j\rangle,
    \label{QFT-psi}
\end{equation}

where $\beta_j$ are the Fourier coefficients, which encode the amplitude and phase information of the transformed state in the Fourier basis. The Fourier coefficients $\beta_j$ are derived from the original coefficients $\alpha_x$ through a linear transformation, as detailed in eq(\ref{qft-fourier}).

    \item Once the QFT has been applied and the Fourier coefficients $\beta_j$ have been obtained, we can extract the phase information associated with each coefficient. The relationship between the Fourier coefficient and its corresponding phase shift can be expressed as:

\begin{equation}
    \beta_j = r_j e^{i\theta_j},
    \label{coefficient}
\end{equation}

where $ r_j = |\beta_j| $ is the magnitude of the coefficient, and $\theta_j$  represents the phase-information associated with the  $j$-th state in the Fourier basis. This phase shift $\theta_j$ carries crucial information about the periodicity and symmetry of the original quantum state.

    \item Analogous to the classical Hilbert transform, which introduces a phase shift of $\pm\pi/2$ to the frequency components of a signal, we define a phase modification in the quantum context based on the nature of these phase shifts \( \theta_j \). The modified phase  $\theta_j'$  is defined as follows:

\begin{equation}
   \theta_j' = 
\begin{cases} 
      \theta_j - \pi/2, & \text{if } \theta_j > 0, \\ 
      \theta_j, & \text{if } \theta_j = 0, \\ 
      \theta_j + \pi/2, & \text{if } \theta_j < 0,
\end{cases}
\label{QFT-sgn}
\end{equation}

where $\theta_j'$ is the phase after it has been shifted based on the initial sign of $\theta_j$. This phase shift is intended to mimic the effect of the Hilbert transform, which systematically adjusts the phase of each component according to its sign, effectively filtering the quantum state.

After applying this phase modulation, we obtain a new quantum state $|\phi'\rangle$, which can be expressed as:

\begin{equation}
    |\phi'\rangle  = \sum^{N-1}_{j=0} \beta'_j |j\rangle,
    \label{eq:phase-modulated}
\end{equation}

where $\beta'_j = r_j e^{i\theta'_j}$ represents the Fourier coefficients with the modified phase $\theta'_j$.

    \item Once we have the phase modulated state in the Fourier basis, we take it back to computation basis to perform the final measurements. This can be done by applying the inverse QFT (IQFT). Mathematically, taking a state $|\phi\rangle$ from Fourier basis to $|\psi\rangle$ in computation basis is represented as, 
    \begin{equation}
        |\psi\rangle = \frac{1}{\sqrt{N}} \sum_{x=0}^{N-1} \left( \sum_{j=0}^{N-1} \beta_j e^{-2\pi i jx / N} \right) |x\rangle,
        \label{inv-QFT}
    \end{equation}
    we can clearly see that if we apply QFT on a state $|\psi\rangle$ and then apply the inverse operation of IQFT, we should get the original state back. However, in the definition of QHT, we change the Fourier coefficients based on phase-shifts. Thus, $\beta_j' \neq \beta_j$. Therefore, the new coefficients, $\alpha_x'$, to states in computational basis can be written as, 

    \begin{equation}
        \alpha_x' =  \frac{1}{\sqrt{N}}\sum_{j=0}^{N-1} \beta_j' e^{-2\pi i jx / N}.
        \label{modulated-alpha}
    \end{equation}

    From eq(\ref{modulated-alpha}), we write the modified quantum state, $|\psi'\rangle$ as, 
    \begin{equation}
        |\psi'\rangle = \sum^{N-1}_{x=0} \alpha_x' |x\rangle.
        \label{eq:psi'}
    \end{equation}
    It's trivial to note that in this new state, the phase of each $\alpha_x'$ is changed based on the changes done in the Fourier basis in step 4. For all $\theta_j>0$, we see an addition of $\pi/2$ phase to respective $\alpha_x'$.
\end{enumerate}
This concludes the steps for the discrete quantum Hilbert transform. The QHT can be used to modify a given quantum state just based on the phase information, by using the Fourier basis for easier phase extraction. 

\section{Examples}
\label{sec:examples}
In this section, we would described some examples of the Hilbert transform, and a graphical representation of the quantum Hilbert transform (QHT) as described in previous sections. 
\subsection{Example of Hilbert transform}
\begin{figure}[h!]
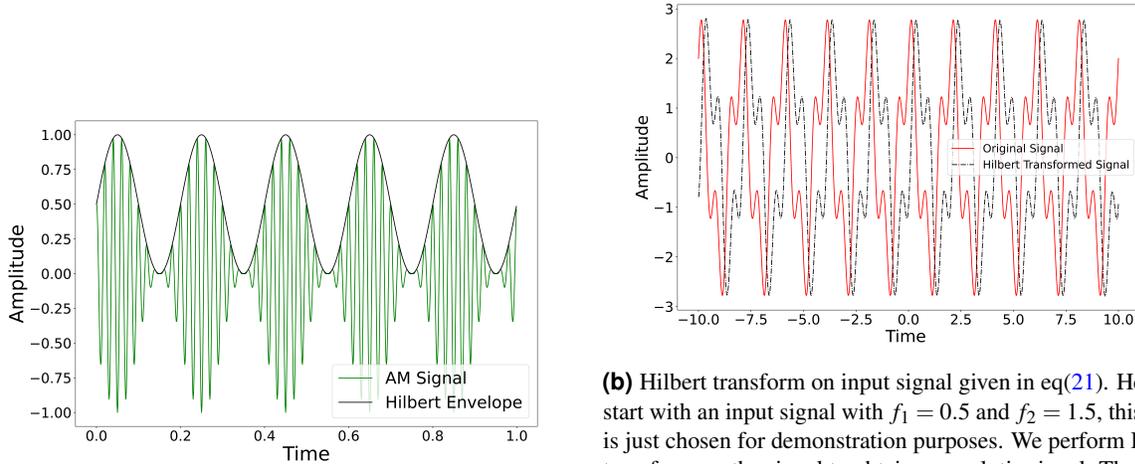

    \centering

    \begin{subfigure}[b]{0.45\linewidth}
        \centering
        \includegraphics[width=\linewidth]{images/amsignal.png}
        \caption{Using Hilbert transform to find the envelope for AM signals. The envelope of the AM signal (shown in blue) is the magnitude of the analytical signal (as shown in red).}
        \label{fig:ams}
    \end{subfigure}    
    \begin{subfigure}[b]{0.45\linewidth}
        \centering
        \includegraphics[width=\linewidth]{images/contiht.png}
        \caption{Hilbert transform on input signal given in eq(\ref{eq:f1}). Here we start with an input signal with $f_1=0.5$ and $f_2=1.5$, this value is just chosen for demonstration purposes. We perform Hilbert transform on the signal to obtain an analytic signal. The red shows the original input signal, and black shows the signal generated after applying Hilbert transform. 
        We can see that the Hilbert transformed signal is phase-shifted by $\pi/2$.}
        \label{fig:HT}
    \end{subfigure}
    
    \caption{(a) Demonstration of using Hilbert transform to find the envelope of AM signals. (b) Application of Hilbert transform to a signal to generate an analytic signal and observe the $\pi/2$ phase shift.}
    \label{fig:combined_HT}
\end{figure}

In this section, we would present some graphical representation of the Hilbert transform over an input signal. We start with a signal as described,
\begin{equation}
    x(t) = 2\cos(2\pi f_1 t)+2\sin(2\pi f_2 t),
    \label{eq:f1}
\end{equation}
where $f_1$ and $f_2$ are the two frequencies. $f_1$ contributes to the cosine component, and $f_2$ contributes to the sine component of the input signal. We perform Hilbert transform on it, and present both the original and transformed signal below. Another important application of Hilbert transform is in broadband modulation. Here, we take a Amplitude Modulated (AM) signal ($x(t)$). We apply Hilbert transform and find the analytical signal ($z(t)$). 
\begin{equation}
    z(t) = x(t) + j\hat{x}(t),
\end{equation}
where $j=\sqrt{-1}$. We can find the magnitude of the analytical signal as, 
\begin{equation}
    |z(t)| = \sqrt{x(t)^2 + \hat{x}(t)^2}
\end{equation}
This gives us the envelope for the signal. It's represented in the fig(\ref{fig:ams}).

\subsection{Example of quantum Hilbert transform}
\begin{figure}[h!]
    \centering
    \begin{subfigure}[b]{0.45\linewidth}
        \centering
        \includegraphics[width=\linewidth]{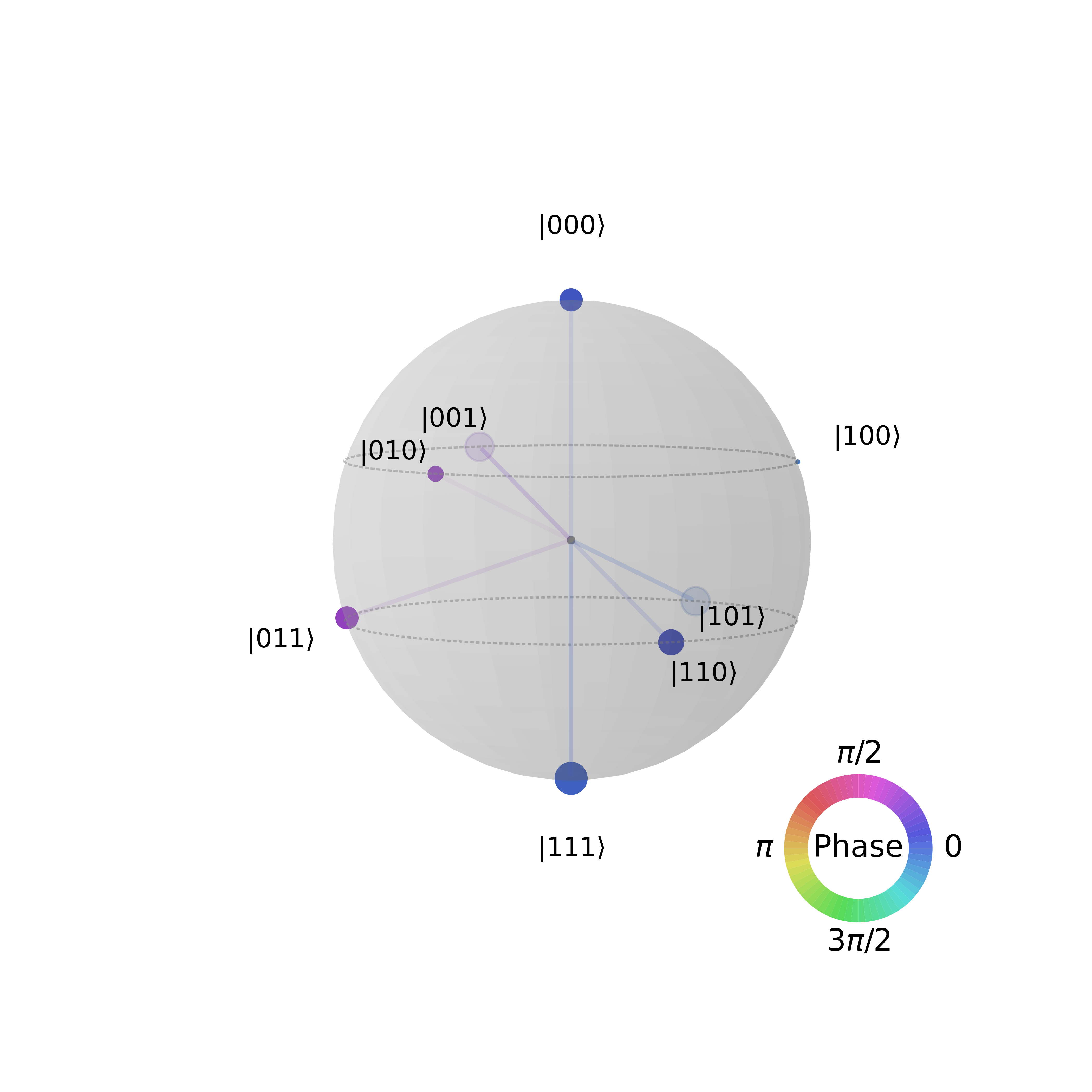}
        \caption{Initial random quantum $3-$qubit state $\ket{\psi}$ with their phase and amplitude magnitude displayed using IBM Q-sphere. Bigger circles on the states represent higher amplitude, and different colors represent the state's phase as represented through the color wheel.}
        \label{fig:initQ}
    \end{subfigure}
    \begin{subfigure}[b]{0.45\linewidth}
        \centering
        \includegraphics[width=\linewidth]{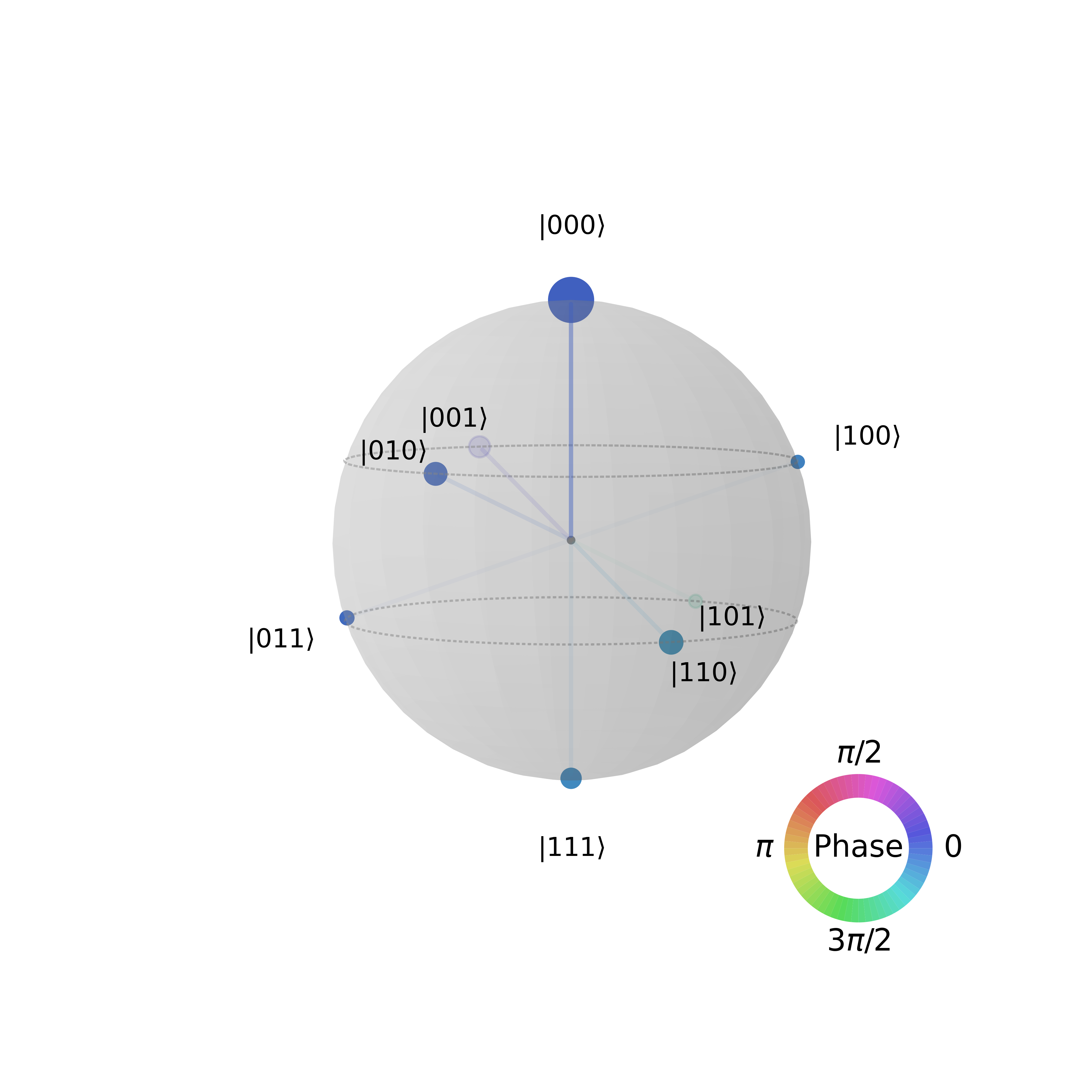}
        \caption{Modified quantum state $\ket{\psi'}$ with their phase and amplitude magnitude displayed using IBM Q-sphere after the application of QHT on the initial quantum state $\ket{\psi}$. Bigger circles on the states represent higher amplitude, and different colors represent the state's phase as represented through the color wheel.}
        \label{fig:finalQ}
    \end{subfigure}
    
    \caption{We plot Q-sphere to represent the multi-qubit state represented by $\ket{\psi}$. We initiate random $\ket{\psi}$ for $N=8$ basis states, i.e., $3$-qubit system, with each state having a random amplitude, all normalized. We then apply QHT as described in Sec[\ref{QHT}], and obtain the transformed state, $\ket{\psi'}$. The states have transformed amplitudes and phases as seen on the right.  }
    \label{fig:qsphere}
\end{figure}

The main focus of this study is to develop a quantum analogue of the classical Hilbert transform for quantum states. We defined the theoretical foundation for QHT in Sec[\ref{QHT}]. The main idea behind QHT is the phase modulation of the quantum states in Fourier Bases. Here, we would apply QHT on some fundamental quantum states (superposition of $2$ qubits, entangled states), and show how the phase is modulated for these. For $N=8$, i.e., $3-$qubit state we plot the initial state $\ket{\psi}$ on Q-sphere and the QHT modified $\ket{\psi'}$ state in fig(\ref{fig:qsphere}).

As explained earlier, due to the rules governing the phase modulation in Fourier basis, the amplitude of the modified quantum state also changed. We would quantify this by plotting the amplitude of the original quantum states vs the amplitude of the states post-QHT. 
\begin{figure}[h!]
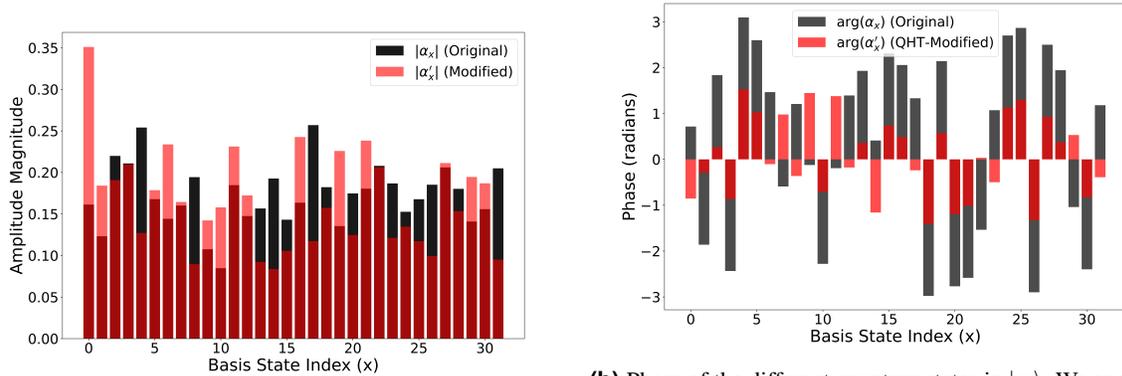

    \centering
    \begin{subfigure}[b]{0.45\linewidth}
        \centering
        \includegraphics[width=\linewidth]{images/amp.png}
        \caption{Amplitude of different quantum states in $|\psi\rangle$ as defined in eq(\ref{quant-state}). The black bars show the original amplitudes, and the red bars show the post-QHT amplitudes.}
        \label{fig:amp}
    \end{subfigure}
    \begin{subfigure}[b]{0.45\linewidth}
        \centering
        \includegraphics[width=\linewidth]{images/phase.png}
        \caption{Phase of the different quantum states in $|\psi\rangle$. We apply QHT to a given quantum state initialized as defined in eq(\ref{quant-state}). We shift the phase of this states based on the signs of the $\theta$ as described in eq(\ref{QFT-sgn}).}
        \label{fig:phase}
    \end{subfigure}
    
    \caption{In this figure base state refers to $|x_i\rangle$ as the quantum state $\ket{\psi}$ is defined as a sequence of basis states defined in eq(\ref{quant-state}) for $N=32$, i.e., $5-$qubit state. (a) Comparison of amplitudes before and after applying QHT. (b) Visualization of phase modifications introduced by QHT.}
    \label{fig:combined_qht}
\end{figure}

\begin{figure}[h!]
    \centering
    \includegraphics[width=0.5\linewidth]{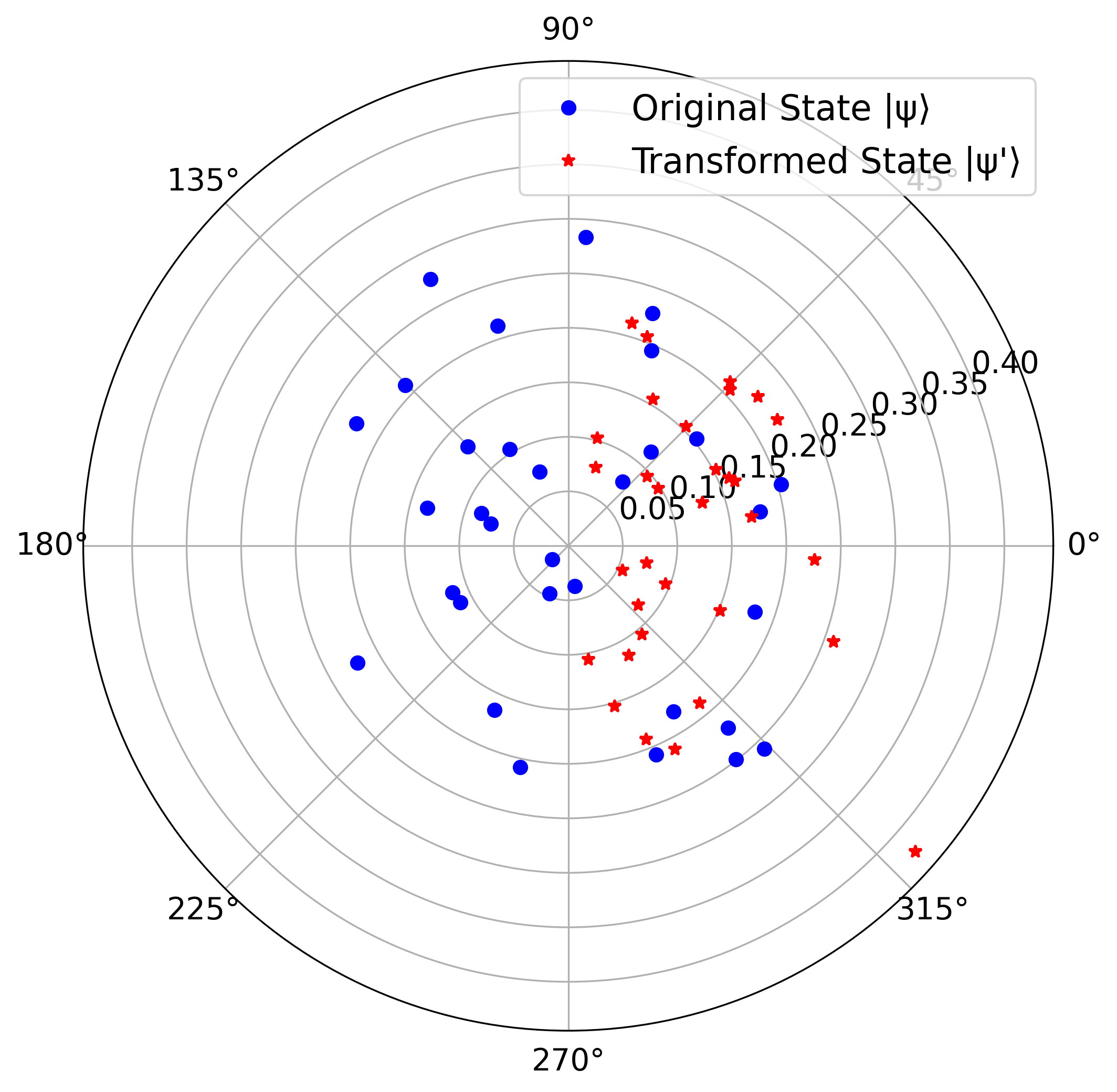}
    \caption{In this figure, red stars show the modified state, and blue dots show the original state. The closer a point is to the origin, the lesser it's amplitude is. The phase is denoted by the angle of the point from the origin.}
    \label{fig:polar}
\end{figure}

We notice from fig(\ref{fig:combined_qht}) that both the amplitude and phase changes for the signal when we apply QHT, due to the nature of the phase modulation. It is also interesting to note that some amplitudes are more suppressed after we apply QHT. We do a polar plot to understand these changes in a simpler manner too in fig(\ref{fig:polar}).

\section{Discussion} 
\label{sec:discussion}
The formulation of QHT allows a user to modulate phase information in quantum states on a Fourier basis. We can expand this idea of phase modulation and hide secret information using QHT. This process is described as follows.

We start with using two quantum states: $\ket{\psi}_\text{ref}$ and $\ket{\psi}_\text{msg}$. The reference state ($\ket{\psi}_\text{ref}$) is used by Bob to get information about the original phase associated with the quantum states, and $\ket{\psi}_\text{msg}$ state is used to encode the secret message. Therefore, this way of sharing a secret message can be conducted in two steps: (1) Transmitting the $\ket{\psi}_\text{ref}$ stage, and Bob obtaining the original phase information. (2) Transmitting the $\ket{\psi}_\text{msg}$ state, which contains the modified phase information using QHT, that is, the hidden message. 
\begin{enumerate}
    \item[Step I:] Transmitting the \textit{ref} state. 
    \begin{itemize}
        \item Alice prepares the quantum state as mentioned in eq(\ref{quant-state}). 
        \item Alice can use a secure protocol to transmit these states to Bob. For example, Alice can use the three-stage protocol\cite{kak2006} to transmit this state. 
        \item After the protocol is completed, Bob can measure the phase information of the quantum states. Bob should now have a list of phase and amplitude information for the reference states. 
    \end{itemize}
    \item[Step II:] Transmitting the \textit{msg} state.
    \begin{itemize}
        \item At this point, Bob already has the list of original phase and amplitude information. Alice creates the same state as Step I. Now, Alice applies QHT to modify the phase information as explained in eq(\ref{QFT-sgn}). 
        \item Alice now adds a sequence at the end of this state to determine which qubits are used to hide secret information, i.e., we only use part of the sequence to hide the secret message. Thus, the length of the secret message ($s$) is less than the original length of the qubit sequence. 

        For example, Alice \textit{ref} state was originally $6$-bits long, and she wants to hide a secret message. Alice would attach a $6$-bit identifier sequence at the end of the original message: $1$ if that qubit contributes to the secret message, $0$ if it does not. At the very end, Alice also puts the encoding state, i.e., $10$ if phase change of $\pi/2$ encrypts $1$, else $01$ if $\pi/2$ encrypts $0$. For this example, the length of the \textit{msg} sequence would be $14$. In general, the length of the \textit{msg} sequence is $l = 2s+2$, where $s$ is the length of the \textit{ref} sequence.      
        
        This entire state is converted into the quantum state. Note that the identifier sequence is only attached after Alice performed QHT on the original sequence. 

        \item Alice would transmit this updated sequence to Bob using the three-stage protocol. Bob performs the following steps, 
        \begin{itemize}
            \item Bob determines the message, identifier, and encryption scheme for the hidden message. 
            \item From the message sequence, Bob identifies the modified phase and amplitudes with respect to the reference state. Bob can thus calculate which qubits had $\pm \pi/2$ phase changes.
            \item Using the identifier sequence, Bob can identify the qubits used to hide the message. Bob can further determine the encryption scheme. 
            \item Using this information, Bob can recover the binary string of hidden information hidden in the modified phase of the quantum states. 
        \end{itemize}
        \begin{tcolorbox}[colback=white,colframe=black,title=Example: Hiding a 3-bit secret in a 6-qubit sequence]
            \begin{itemize}[leftmargin=*]
              \item \textbf{Reference bits (6):} 
                \[
                  1\;0\;0\;1\;0\;1
                  \quad\bigl(\ket{1}\ket{0}\ket{0}\ket{1}\ket{0}\ket{1}\bigr)
                \]
              \item \textbf{Secret message bits (3):}
                \[
                  1\;0\;1
                \]
              \item \textbf{Identifier sequence (6):}
                \[
                  1\;0\;1\;0\;1\;0
                  \quad\text{(marks qubits 1,\,3,\,5 for hiding)}
                \]
              \item \textbf{Encoding scheme bits (2):}
                \[
                  10
                  \quad\bigl(\tfrac{\pi}{2}\text{ shift}\to\texttt{“1”}\bigr)
                \]
               \item \textbf{Combined $\ket{\psi}_{\text{msg}}$ (14 qubits):}
                \[
                  \ket{\psi_{\text{msg}}}
                  = \fbox{$\ket{100101}'$}
                    \;\fbox{$\ket{101010}$}
                    \;\fbox{$\ket{10}$}
                \]
                where $\ket{100101}'$ refers to the updated reference state due to application of QHT.
            \end{itemize}
        \end{tcolorbox}

    \end{itemize}
Here we presented an application of the formulation of the QHT we described in this work, where Alice can hide some secret information in the phase of the quantum states transmitted using QHT and steps outlined above.

\end{enumerate}

This method utilities a quantum communication protocol to transmit a hidden message embedded using phase modification in qubit states. Since, it needs Alice to pre-share a reference state, an older communication between Alice and Bob can be used to eliminate the need of sharing the reference state again. Alice and Bob can, thus, score phase information for a shared communication and use it for future use as the reference state. Moreover, the receiver needs to know the exact qubit index whose phase were modified to extract the secret message, it adds a layer of uncertainty for the eavesdropper. As a result, to decode the secret message eavesdropper needs three quantities: (1) original phase information from reference state (2) encoding detail and qubit indices (3) decryption for communication protocol. The information leakage can, thus, be approximated based on the communication protocol used. For example, we proposed using three-stage protocol to conduct this. So the information leakage can be quantified by Holevo bound\cite{hall1997quantum, holevo1973bounds} for three-stage protocol. 

\begin{enumerate}
    \item For noise-less (i.e., ideal case), an eavesdropper intercepting one of the three stages of communication gets a maximally mixed state, the reduced density matrix can be written as,
    \begin{equation}
        \rho_E = \frac{1}{N}\left( \sum_{x=0}^{N-1}\ket{x}\bra{x} \right),
        \label{eq:mixed}
    \end{equation}
    which is independent of $\ket{\psi}_\text{ref}$ or $\ket{\psi}_\text{msg}$ state. As per, Holevo's bound the mutual information accessible to eavesdropper for $s$-qubit payload, let's say $M$, can be written as,

    \begin{equation}
      \chi 
  = S\left(\frac{1}{N}\sum_m \rho_m\right)\;-\;\frac{1}{N}\sum_m S(\rho_m)
  = \log N \;-\;\log N
  = 0,    
    \end{equation}
    Therefore, for ideal case, $I_{\text{Eve}|M}=0$.

    \item For practical case, where the gate and channel noises ($\delta$) disrupts the perfect mixing, and introduces a small information leakage that can be bounded as follows, 

    \begin{equation}
        I(\mathrm{Eve}:M)
      \;\le\;
      H_2\!\left(\tfrac{1 + \sqrt{1-\delta}}{2}\right)
      \;\approx\;
      H_2\!\left(1 - \tfrac{\delta}{4}\right)
      \;\sim\; O(\delta),
      \label{eq:mutual-noise}
    \end{equation}
    where $H_2(p) = -p\log_2 p - (1-p)\log_2(1-p)$ is the binary entropy function. Therefore, if $\delta\ll 1$, the information leakage can be minimized. 
\end{enumerate}

\section{Conclusions}
\label{Conclusions}

This work presents a quantum analogue of the classical Hilbert transform for quantum states. We explain the basics of classical Hilbert transform (both continuous and discrete), go over the definitions of Fourier and quantum Fourier transform and present the proposed working of quantum Hilbert transform (QHT). We also present a use case for QHT where a sender can use QHT to hide information in a particular quantum state, which can be retrieved by performing a phase analysis.

\bibliography{main}

\section*{Authors Contributions}
N.J. is the primary author of the manuscript and received intellectual inputs from A.P.

\section*{Data availability statement}
All data generated or analyzed during this study are included in this published article [and its supplementary information files].

\section*{Additional Information (Competing Interests)}
None

\end{document}